# Tunable interplay between light and heavy electrons in twisted trilayer graphene

Andrew T. Pierce[1*‡#], Yonglong Xie[1,2,3*‡], Jeong Min Park[2*], Zhuozhen Cai[1], Kenji Watanabe[4], Takashi Taniguchi[5], Pablo Jarillo-Herrero[2‡], Amir Yacoby[1‡]

[1]Department of Physics, Harvard University, Cambridge, MA 02138, USA
[2]Department of Physics, Massachusetts Institute of Technology, Cambridge, MA 02139, USA
[3]Department of Physics and Astronomy, Rice University, Houston, TX 77005
[4]Research Center for Electronic and Optical Materials, National Institute for Material Science, 1-1 Namiki, Tsukuba 305-0044, Japan
[5]Research Center for Materials Nanoarchitectonics, National Institute for Material Science, 1-1 Namiki, Tsukuba 305-0044, Japan

[*]These authors contributed equally to this work.
[‡]Corresponding authors' emails: atp66@cornell.edu, yx71@rice.edu, pjarillo@mit.edu, yacoby@g.harvard.edu
[#]Present address: Kavli Institute at Cornell for Nanoscale Science, Ithaca, NY, USA

**Abstract:** In strongly interacting systems with multiple energy bands, the interplay between electrons with different effective masses and the enlarged Hilbert space drives intricate correlated phenomena that do not occur in single-band systems[1,2]. Recently, magic-angle twisted trilayer graphene[3–7] (MATTG) has emerged as a promising tunable platform for such investigations: the system hosts both slowly dispersing, "heavy" electrons inhabiting its flat bands as well as delocalized "light" bands that disperse as free Dirac fermions[8]. Most remarkably, superconductivity in twisted trilayer graphene[3–7] and multilayer analogues with additional dispersive bands[9–11] exhibits Pauli limit violation and spans a wider range of phase space compared to that in twisted bilayer graphene, where the dispersive bands are absent. This suggests that the interactions between different bands may play a fundamental role in stabilizing correlated phases in twisted graphene multilayers. Here, we elucidate the interplay between the light and heavy electrons in MATTG as a function of doping and magnetic field by performing local compressibility measurements with a scanning single-electron-transistor microscope. We establish that commonly observed resistive features near moiré band fillings $\nu=-2, 1, 2$ and $3$ host a finite population of light Dirac electrons at the Fermi level despite a gap opening in the flat band sector. At higher magnetic field and near charge neutrality, we discover a new type of phase transition sequence that is robust over nearly 10 micrometers but exhibits complex spatial dependence. Mean-field calculations establish that these transitions arise from the competing population of the two subsystems and that the Dirac sector can be viewed as a new flavor analogous to the spin and valley degrees of freedom. These findings shed new light on the role of dispersive bands and establish twisted graphene multilayers as a platform for realizing correlated multiband ground states.

**Main text**

In magic-angle twisted bilayer graphene (MATBG), the flat bands in the presence of strong interactions lead to novel many-body phenomena, including unconventional superconductivity[1,2] and correlated incompressible states[13] with nontrivial topology[14,15]. To understand the superconducting state, early theory efforts[16–21] primarily focused on models that emphasize the heavy electrons in the isolated flat bands. The experimental discovery of robust superconductivity in magic-angle twisted trilayer graphene (MATTG)[3,4] —whose band structure at zero displacement field contains a set of highly dispersive bands in addition to the flat bands[8]—suggests that the addition of dispersive electrons may be important for stabilizing superconductivity in this system. This conjecture is made even more enticing by the recent observation of superconductivity in magic-angle twisted quadri- and pentalayer graphene[9–11] whose band structures contain two or more dispersive bands. These developments highlight the universal, non-trivial role of dispersive electrons in determining the ground states in twisted graphene multilayers and suggest that the interplay between the light and heavy electrons may be key to understanding superconductivity in these systems.

To elucidate the nature of the coupling between the light and heavy electrons in graphene multilayers, we present high-resolution local compressibility measurements on MATTG obtained using a scanning single-electron-transistor (SET) microscope. We chose to study MATTG because of the simplicity of its band structure (Fig. 1a), which consists of flat bands nearly identical to those of MATBG along with a massless Dirac-like band nearly identical to that of monolayer graphene, with the hybridization between these two subsystems being tunable by means of an applied displacement field. We observe for the first time robust incompressible states, with gaps of several meV, that remain stable but exhibit sudden density shifts induced by the presence of dispersive Dirac electrons as the magnetic field is reduced to zero. Using the SET as a local discrete charge sensor, we establish that these zero-field incompressible peaks correspond to many-body ground states in which a gap is opened only in the flat band sector while the density of dispersive electrons remains finite. This provides a clear explanation for the enigmatic resistive peaks that display metallic behavior observed in transport measurements. While we conclude that the zero-field many-body states near integer fillings can be understood to be a consequence of the weak coupling between the Dirac and flat band electrons, we also discover a new type of phase transition near charge neutrality and at high magnetic field that instead results from strong coupling between the two subsystems. Taken together, our results shed new light on the role of dispersive electrons in determining the correlated ground states, including superconductivity, in twisted graphene multilayers.

The device we study in this work is a single-gated hexagonal boron nitride (hBN)-encapsulated MATTG device (see Methods). Transport measurements verify that the device exhibits the hallmarks of interaction-driven phenomena in MATTG, including robust superconductivity near $\nu=\pm2$ (Extended Data Fig. 1). Local compressibility measurements were obtained with the SET using a protocol described in detail elsewhere[22]. While a finite displacement may be present in this device geometry, we show below that our data are consistent with negligible hybridization between Dirac and flat bands.

**Coexistence of light and heavy electrons**

We begin by showing that the general characteristics of the ground states in our single-gated MATTG device resemble those of MATBG but with a superimposed massless Dirac band. Fig. 1b shows local inverse compressibility dµ/dn as a function of out-of-plane magnetic field $B$ and moiré band filling factor $v$. Our measurements reveal a rich phase diagram, including incompressible states exhibiting peaks in dµ/dn whose Chern numbers $C$ can be extracted from their trajectories in magnetic field using the Streda formula[23] $dn/dB=C\phi/\phi_0+s$. The principal Chern insulators observed in the experiment are tabulated in the Wannier diagram of Fig. 1c. Consistent with previous experiments[3,4,7], we observe larger Chern numbers at partial flat band fillings in MATTG than can occur in MATBG, signaling the presence of the light Dirac band coexisting with the heavy flat bands. This can be understood by separating the contributions from the two subsystems using $C=C_d + C_f$, where $C_d$ and $C_f$ are the Chern numbers of the filled light Dirac and heavy flat bands respectively. Specifically, while the heavy flat bands are only weakly perturbed by an applied magnetic field, as little as 0.5 T is enough to break the light Dirac band into widely gapped fourfold-degenerate Landau levels (LLs), with the $N=0$ LL remaining exactly at the energy of the $B=0$ Dirac point (Fig. 1d). These considerations allow us to isolate $C_f$, by subtracting +2 or −2 from the total Chern number $C$ if the Fermi level lies in the conduction or valence flat bands respectively. The observed sequence of $C_f$ in MATTG is in full accord with the known phenomenology in MATBG, including states emanating from integer and half integer fillings (Extended Data Fig. 2) that break the translation symmetry of the system[24]. Further signatures of the additional light Dirac bands come from the experimentally determined many-body Hofstadter spectrum (Fig. 1f). While the general features of the spectrum resemble that of MATBG[25], we observe a series of rapidly dispersing features with enhanced density of states (red dots in Fig. 1f), which we attribute to the $N\neq 0$ Dirac LLs. Careful analysis of the energy separation between the neighboring levels yields a monolayer-graphene Dirac velocity $v_F\sim 10^6$ m/s (see Methods), in agreement with theoretical expectations. These general characteristics have been confirmed at multiple locations within the device and are robust against changes in the local twist angle (see Extended Data Figs. 2 and Methods). Overall, the good correspondence between states observed in MATTG and MATBG confirms our interpretation of the high-field portion of the phase diagram in terms of the contributions of the light and heavy bands and suggests that the hybridization between the two subsystems is negligible.

**Partially gapped states at zero magnetic field**

Having established the coexistence of light and heavy bands in MATTG, we now focus on the low magnetic field and finite filling factor portion of the phase diagram, where the light and heavy carriers conspire to form various many-body ground states including superconductivity. Strikingly, the incompressible peaks emerging from $v=1$, $\pm 2$, and 3 appear stable down to zero magnetic field (Fig. 2a-d), despite deviating from the high-field linear trajectories at low field $B$<0.5 T. In the same magnetic field range, rapidly dispersing *compressible* features appear to cut through the incompressible peaks. While similar features were observed in previous transport measurements and identified as the $|N|$>0 Landau levels of the light Dirac bands[3,4,7], these experiments did not find any zero-magnetic-field incompressible states at zero or small displacement field.

An important characterization of these new and unexpected incompressible states comes from the analysis of the density at which they emerge at zero magnetic field, as this enables us to determine the relative fillings $\nu_f$ and $\nu_d$ of the flat and dispersive bands that satisfy $\nu = \nu_f + \nu_d$. First, we examine the high-field trajectories of the incompressible states emanating from the partially filled flat bands. Because the Dirac sector is fully gapped at moderate magnetic field (~0.5T), it follows that the $\nu$-intercepts to which the high-field trajectories of the incompressible states extrapolate at $B$=0 occur where $\nu_f$ is an integer. The observation that the incompressible peaks at $B$=0 are shifted from integer $\nu$ therefore implies that $\nu_d$ is non-zero when the system becomes incompressible (see Methods). We therefore conclude that a finite number of light Dirac electrons must participate in the formation of these unexpected incompressible states. While the flat bands of MATTG may be gapped out by spontaneous symmetry breaking in analogy to correlated incompressible states observed in MATBG, the Dirac point of the unbounded monolayer graphene-like Dirac band is more robust against these symmetry-breaking effects, hindering the occurrence of true insulators—i.e., states with zero density of states at the Fermi level—at zero magnetic field.

One fundamental distinction between true insulators and systems with small but nonzero density of states is their response to an underlying disorder landscape, which can be discerned using our ultra-local electrometry measurements. For any given disorder potential (Fig. 2e), if the system is compressible, the resulting self-consistent density profile reconfigures itself to cancel the disorder potential (Fig. 2f, upper panel). However, as the system becomes more and more insulating, it breaks apart into a network of localized, weakly-connected compressible regions (Fig. 2f, lower panel). Eventually, the remaining compressible regions become so small that they can themselves exhibit Coulomb blockade, yielding a strong electrostatic signature as electrons are added one-by-one into the puddles. Consequently, the incompressible peaks exhibited by a true insulator with zero density of states are comprised of many discrete peaks, with each peak corresponding to charging a separate small region of the system. Moreover, the spatial distribution of these quantum-dot-like regions—known as localized states—provides additional information about their local electrostatic environments, which can be tuned by applying a voltage bias between the tip and the sample. By contrast with a true insulator, a system with low but nonzero density of states retains its ability to screen the local variation of the electrostatic potential, resulting in the absence of localized states.

By operating the scanning SET as a local discrete charge sensor (see Methods), we resolve the fine structure of the incompressible peaks observed at $\nu$=+1, ±2, and +3 (Fig. 2g-j). We observe that, at high magnetic field, the incompressible peaks are indeed composed of many discrete peaks, each corresponding to the addition of a single electron. Spatial measurements further confirm that these peaks correspond to adding electrons to discrete, local charge puddles, the profile of which can be controlled by the voltage difference between the tip and the sample as expected for localized states (see Extended Data Fig. 3). Strikingly, however, as the magnetic field is reduced, the discrete peaks vanish below a critical value of magnetic field. In the case of the incompressible peak from $v=+2$, the discrete peaks are seen to re-enter the phase diagram briefly before finally disappearing as the field approaches zero. We attribute the disappearance of the localized states to the presence of a finite density of Dirac fermions at the Fermi level at very low to zero magnetic field, which screens the electrostatic disorder landscape and therefore destroys the discrete charging peaks. Thus, the robust zero-field incompressible peaks, together with the absence of the localized states, enable us to establish that the light Dirac bands remain populated within the integer-filling flat band gaps independent of any particular microscopic model.

**Unexpected phase transitions near charge neutrality**

While the phases at finite filling factors are well described by the weakly-coupled Dirac and flat bands, high-resolution measurements near charge neutrality at high magnetic field reveal signatures of strong interactions between the electrons from these two subsystems. Fig. 3a depicts the local compressibility measured near charge neutrality in terms of LL filling $v_{LL}=nh/eB$, where $n$ is the electron density. The observation that the strongest incompressible peaks occur at $v_{LL}=\pm 6$ verifies the expected presence of the four-fold Dirac $N=0$ LL and the eight-fold $N=0$ LL from the flat bands near charge neutrality. Moreover, we resolve incompressible peaks at each integer filling between $v_{LL}=\pm 6$, signaling that strong interactions fully lift the degeneracy of the LLs in both the Dirac and flat band sectors. Most notably, four pronounced regions of giant negative compressibility interrupt the incompressible peaks at $v_{LL}=-5$, $-4$, $-3$, $-2$, $-1$ and $0$, and surprisingly cross through regions that are otherwise completely compressible. We note that a sequence of additional peaks away from integer fillings is also apparent; these weaker states arise due to the supermoiré pattern in the system[26] and are not expected to play an important role in the emergence of strong negative compressibility.

The interruptions of the incompressible peaks by regions of sharp negative compressibility strongly suggest phase transitions in which the isospin polarization and the occupation of the subsystems abruptly changes. Previously, similar features have only been observed in the fractional quantum Hall regime and were ascribed to isospin transitions of composite fermion LLs[27]. In that experiment, the number of transitions at a given fractional quantum Hall state directly reflects the available degrees of freedom in the system and the magnetic fields at which the incompressible peaks are interrupted correspond to the crossings of the composite fermion LLs that are dictated by their energy separations and effective $g$-factors. The appearance of sharp phase transitions in the integer quantum Hall regime is highly unusual because the limited number of available degrees of freedom in the integer quantum Hall system normally gives rise to tightly constrained phase diagrams, such as that of the $v=1$ quantum Hall ferromagnetic state. Therefore, the observed phase transitions point towards a rich array of ordered electronic states, enabled by the high degeneracy of Dirac and flat band LLs and the strong correlation between them. Moreover, the fourfold multiplicity suggests a connection to the fourfold-degenerate $N=0$ Dirac LL.

To verify this picture, we consider the single-particle energy spectrum of the system very close to charge neutrality. Fig. 3b shows two sets of LLs emanating from the Dirac points of the Dirac and flat bands, that are energetically separated by $E_{\text{offset}}$, as established by single-particle band structure calculations[8]. We assume that each LL disperses linearly with magnetic field with a flavor-dependent $g$-factor, resulting in a sequence of level crossings between Dirac and flat band LLs. Despite its simplicity, the energy spectrum in Fig. 3b not only satisfactorily reproduces the magnetic fields at which the incompressible peaks are interrupted, but also predicts energy gaps whose evolution qualitatively matches those we find experimentally (Extended Data Fig. 4). The good agreement confirms our hypothesis that each interruption of the incompressible peaks corresponds to the depopulation of one Dirac LL and places specific constraints on the single-particle band structure as well as the underlying isospin structure of the Landau levels (see Methods).

The unusual observation of extended regions of negative compressibility that are "detached" from the incompressible states can be captured by a model in which the energy levels in Fig. 3b are coupled by an on-site Coulomb potential $U$, similar to those that have previously been employed to describe isospin transitions in MATBG[28,29]. The inverse compressibility can be computed as a function of magnetic field by minimizing the free energy at each value of $v_{\text{LL}}$ (see Methods) and is shown in Fig. 3c. We find that an onsite interaction $U\sim 20W$, where $W$ is the LL width, is sufficient to qualitatively reproduce the experimental data, including the peculiar trajectories of the negative compressibility features as a function of magnetic field. Closer inspection of the filling factors in the Dirac and flat band sectors (Fig. 3d) demonstrates that the negative compressibility marks the phase boundaries at which the population of Dirac LLs, $n_D$, abruptly jumps and enables us to sketch a phase diagram at *all* LL fillings according to $n_D$. The remarkable correspondence between our data and the model calculation highlights the complex interplay between the electrons in the Dirac and flat band sectors and suggests that the Dirac sector can be viewed as a new quantum degeneracy in the Hilbert space analogous to the spin and valley degrees of freedom.

Lastly, we present spatial compressibility measurements over an extended range of 9 μm (Fig. 4a), highlighting the remarkable reproducibility of the phase transitions near the CNP. However, the filling factors at which these transitions appear, $v_i$, exhibit spatial variation (Fig. 4b), suggesting that the effective $g$-factors vary spatially. This is confirmed by examining the magnetic field dependence of the negative compressibility features at a second location (Extended Data Fig. 5a), where they evolve in a manner broadly similar to the measurements presented in Fig. 3, but produce different fit results for their $g$-factors (Extended Data Fig. 5b and Methods). To gain insight into the origin of these variations, we determine the local twist angle based on the quantum

Hall states emanating from the full fillings of the moiré unit cell, as depicted in Fig. 4c. Our analysis suggests that the twist angle variations are not solely responsible for the variations in $v_i$: while $\theta$ and $v_i$ exhibit similar trends for positions within the range of positions from 0 to 5 µm, the rapid evolution of $\theta$ for positions beyond 5 µm is not reflected in the $v_i$, which remain virtually unchanged. These observations indicate that microscopic parameters other than the twist angle contribute to determining the flavor-dependent $g$-factors and highlight the power of the observed phase transitions as a local probe of the electronic environment. Further theoretical and experimental work will be required to fully understand both the microscopic origin of the observed spatial variation and how this may be exploited to control the isospin structure of the system.

The experiments presented here illuminate the intricate relationship between light and heavy electrons and establish clear guidelines for models of the many-body ground states in MATTG, including superconductivity. A key unresolved question is the fate of the excess light Dirac electrons upon increasing doping and whether they play any role in stabilizing superconductivity. Additionally, the correlation between coexisting light and heavy electrons reported here motivates further experimental efforts to investigate Kondo physics and other strongly correlated phenomena in twisted graphene systems, as proposed by recent theoretical advancements[30,31]. Notably, it was recently proposed that MATBG can be mapped to a topological heavy fermion model in which light and heavy electrons strongly interact to form a plethora of correlated quantum phases[32–36].

**Acknowledgements**

We acknowledge discussions with Ady Stern. We thank Ashvin Vishwanath, Eslam Khalaf, Daniel Parker, Patrick Ledwith and Jie Wang for collaboration on related projects. This work was sponsored by the Army Research Office under award number W911NF-21-2-0147 and by the Gordon and Betty Moore Foundation EPiQS initiative through Grant GBMF 9468. Help with transport measurements and data analysis were supported by the National Science Foundation (DMR-1809802), and the STC Center for Integrated Quantum Materials (NSF Grant No. DMR-1231319). P.J-H acknowledges support from the Gordon and Betty Moore Foundation's EPiQS Initiative through Grant GBMF9463. A.T.P. acknowledges support from the Department of Defense through the National Defense Science and Engineering Graduate Fellowship (NDSEG) Program. Y.X. acknowledges partial support from the Harvard Quantum Initiative in Science and Engineering. Y.X, A.T.P. and A.Y. acknowledge support from the Harvard Quantum Initiative Seed Fund. K.W. and T.T. acknowledge support from the JSPS KAKENHI (Grant Numbers 21H05233 and 23H02052) and World Premier International Research Center Initiative (WPI), MEXT, Japan. This work was performed, in part, at the Center for Nanoscale Systems (CNS), a member of the National Nanotechnology Infrastructure Network, which is supported by the NSF under award no. ECS-0335765. CNS is part of Harvard University.

**Author Contributions**

A.T.P., Y.X., J.M.P., P.J.-H. and A.Y. designed the experiment. A.T.P., Y.X. and Z.C. performed the scanning SET experiment, the temperature-dependent transport measurements and analyzed the data with input from A.Y. J.M.P. and P.J.-H. designed and provided the samples and contributed to the analysis of the results. Y.X, A.T.P and Z. C. carried out the simulation of the compressibility. K.W. and T.T. provided hBN crystals. All authors participated in discussions and in writing of the manuscript.

**Competing interests**

The authors declare no competing interests.

**Figure 1**

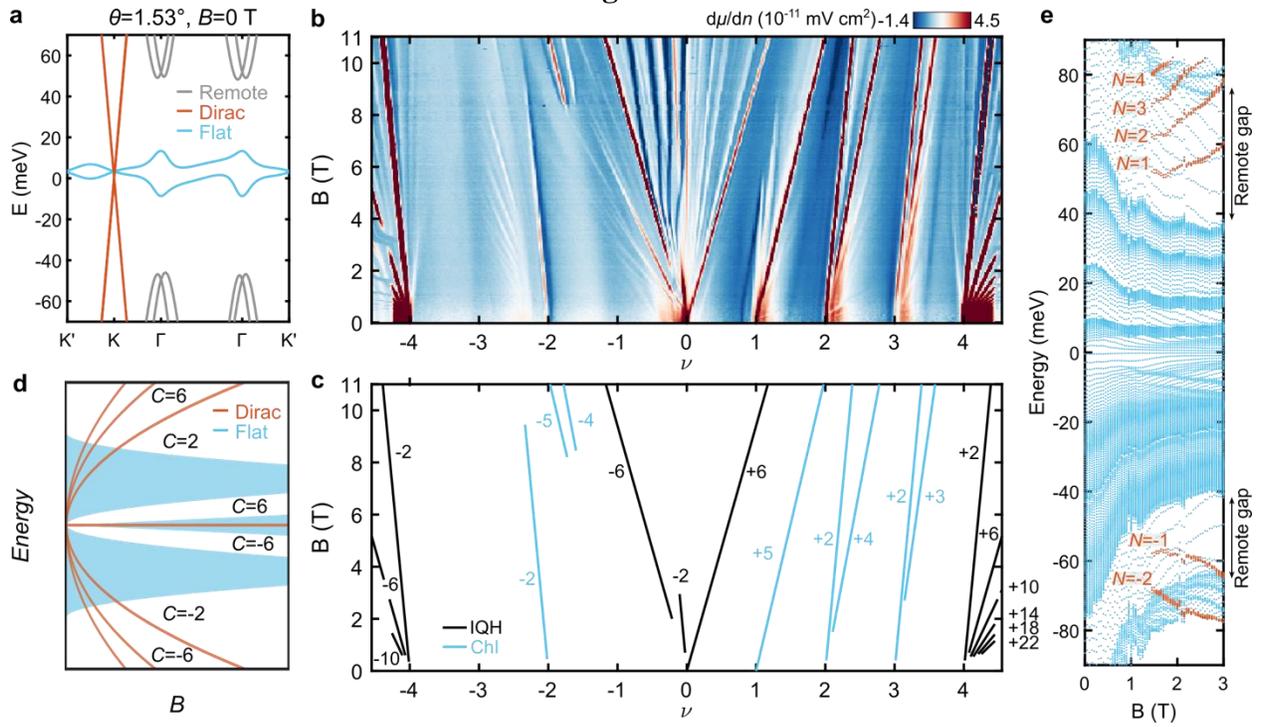

**Figure 1. | Correlated incompressible states from the unhybridized light and heavy bands of MATTG. a**, Zero-field band structure of MATTG with a twist angle of 1.53° in the absence of a displacement field exhibiting the heavy flat bands (blue), the light "Dirac-like" bands (orange), and the remote bands (grey). **b**, Local inverse compressibility dµ/dn plotted as a function of electron density, in units of electrons per moiré unit cell ν, and magnetic field B. **c**, Wannier diagram highlighting the incompressible states with the largest gaps observed at each filling factor. The solid lines indicate the range of magnetic field where each state remains gapped and are labeled by the state's Chern number. **d**, Qualitative sketch of the energy spectrum of MATTG in a finite magnetic field and in the absence of a displacement field showing the expected dispersion of the light and heavy bands (the remote bands are omitted for simplicity). **e**, Experimentally determined Hofstadter energy spectrum showing the light and heavy bands in agreement with Fig. 1d, along with the remote gaps (see main text and Methods section).

**Figure 2**

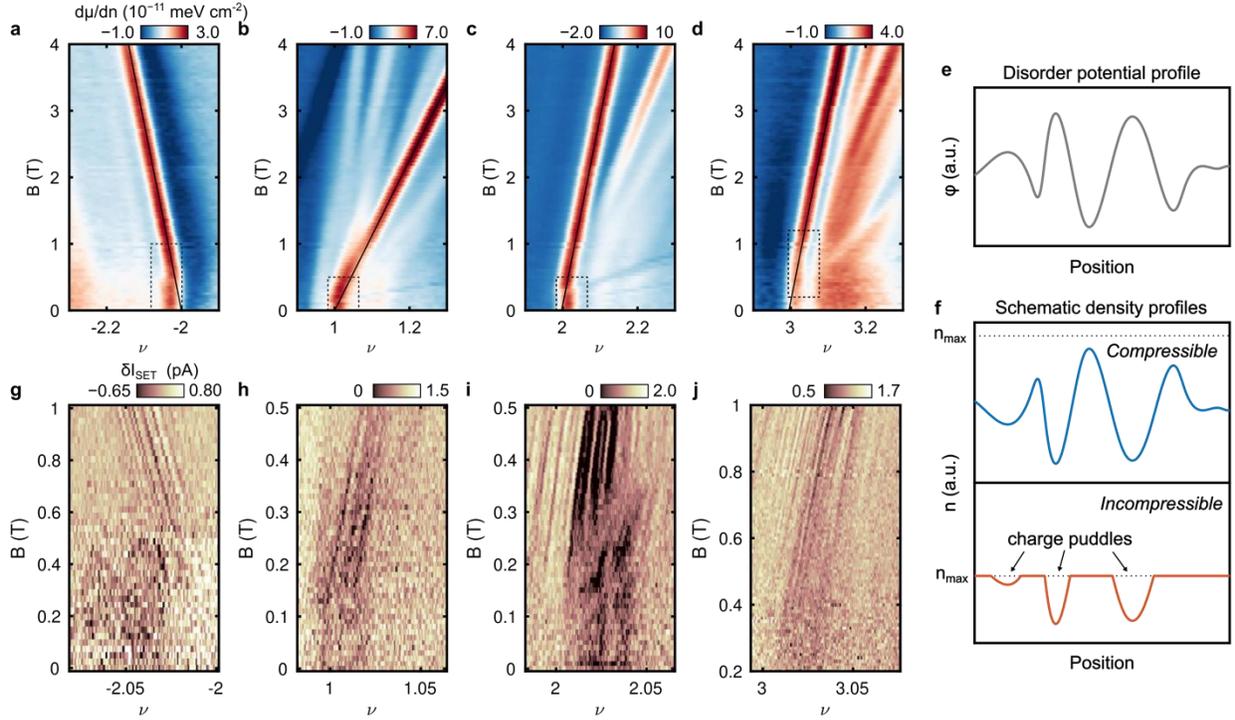

**Figure 2. | Fingerprints of light Dirac fermions spanning the correlated heavy band gaps. a-d,** zoom-in of inverse compressibility at low field near the principal incompressible states at partial fillings of the bands. The black lines are drawn to extrapolate the slope of the high-field Chern insulators back to zero magnetic field; their intercepts on the ν axis determine the exact integer filling of the flat bands. The dashed boxes mark the magnetic field and density ranges in which the data in panel **f-i** were taken. **e,** Schematic disorder potential profile. **f,** Upper: when the sample is compressible, the resulting density profile takes the opposite shape to the intrinsic disorder landscape shown in **e**. Lower: when the sample is insulating, the density at most of the regions is pinned to $n_{max}$ due to the presence of a true energy gap. The remaining areas form charge puddles exhibiting localized states, whose energy levels are determined by the charging energy of the puddles. **g-j**, high resolution penetration electric field measurements over narrow density ranges and at low field in the vicinity of the incompressible peaks. The incompressible peaks can be resolved into localized states at high field, but the localized states disappear at low field when the Dirac bands approach the Fermi level.

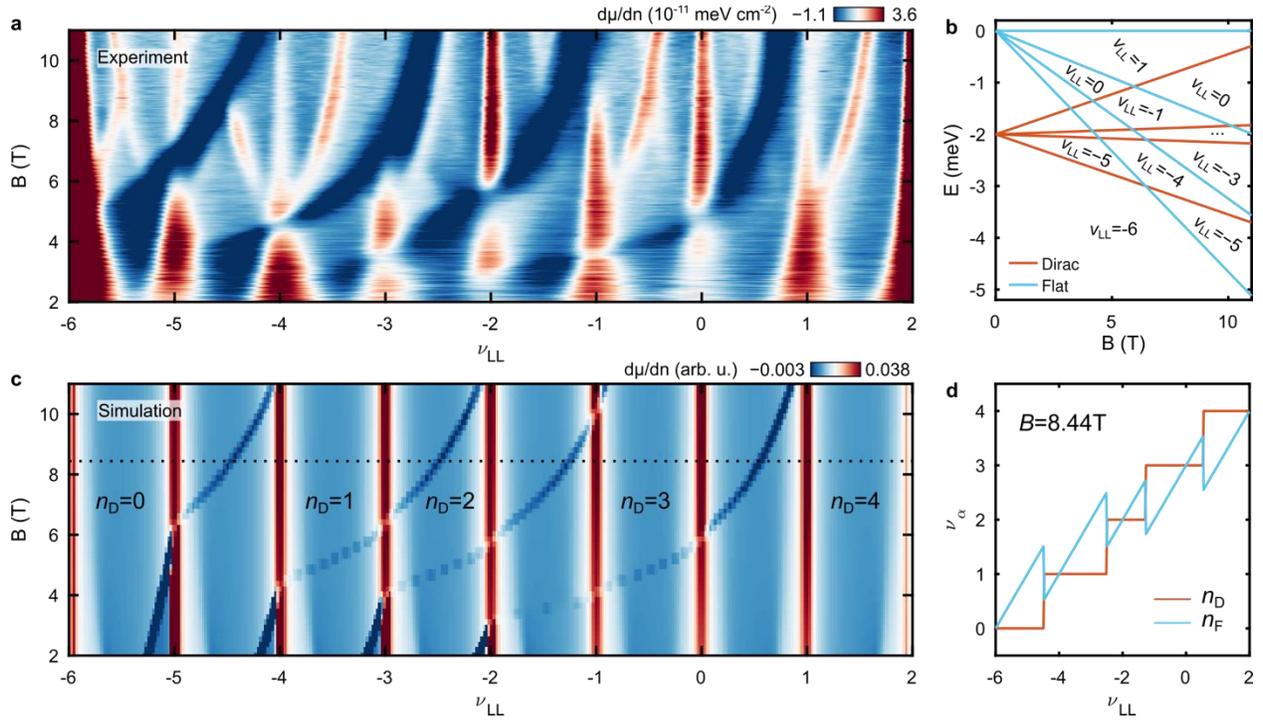

**Figure 3. | Phase transitions near charge neutrality due to strong coupling between Dirac and flat band sectors. a,** Local compressibility of MATTG in the vicinity of charge neutrality plotted as a function of Landau level filling factor $\nu_{LL}$ and magnetic field. **b,** Level diagram showing the dispersion of the four lowest flat band levels and the four Dirac levels as determined by the Landau level crossing points in **a**. **c,** Simulated inverse compressibility using the mean-field model described in the text (see also Methods section). $n_D$ represents the number of Landau levels that are partially or fully populated with carriers.

**Figure 4**

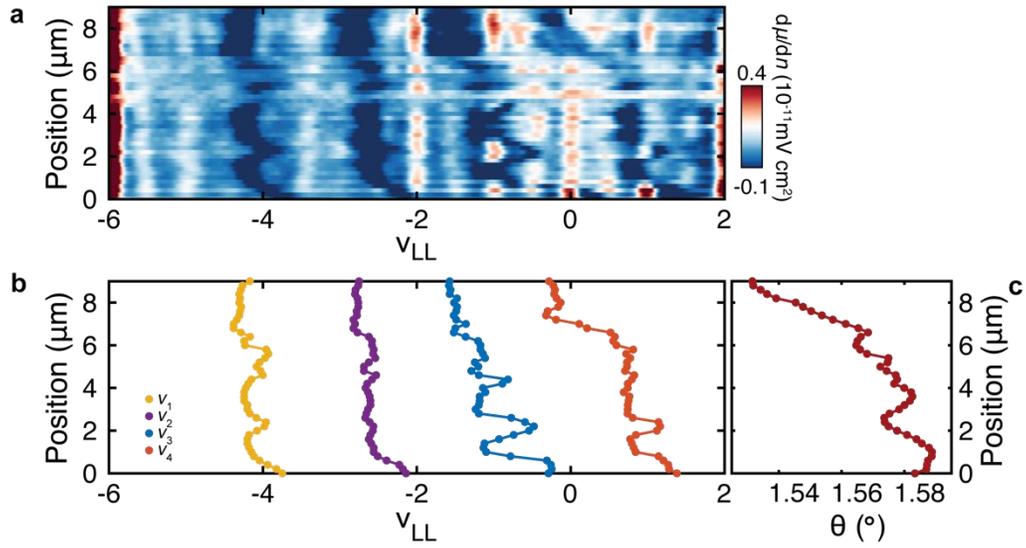

**Figure 4. | Phase transitions near charge neutrality observed over an extended range of 9 μm. a,** Spatial dependence of local inverse compressibility dμ/dn as a function of electron density measured at $B$=11 T **b,** Trajectories of the filling factors $v_i$ at which the phase transitions appear in **a**. **c**, Local twist angle as a function of position.